\documentclass[preprint,aps,12pt,preprintnumbers,eqsecnum,nofootinbib]{revtex4}
\usepackage{graphicx,slashed}
\usepackage{subfigure}

\usepackage{color}
\usepackage{amssymb,amsmath}
\usepackage{epstopdf}

\newcommand{\beq}{\begin{equation}}
\newcommand{\enq}{\end{equation}}
\newcommand{\be}{\begin{equation}}
\newcommand{\ee}{\end{equation}}
\newcommand{\bea}{\begin{eqnarray}}
\newcommand{\eea}{\end{eqnarray}}
\newcommand{\nn}{\nonumber}
\newcommand{\Pl}{\rm{P}}

\begin{document}
%
\title{\vspace*{0.5in} 
Emergent Gravity from Vanishing Energy-Momentum Tensor
\vskip 0.1in}
\author{Christopher D. Carone}\email[]{cdcaro@wm.edu}\affiliation{High Energy Theory Group, Department of Physics,
College of William and Mary, Williamsburg, VA 23187-8795, USA}
\author{Joshua Erlich}\email[]{jxerli@wm.edu}\affiliation{High Energy Theory Group, Department of Physics,
College of William and Mary, Williamsburg, VA 23187-8795, USA}
\author{Diana Vaman}\email[]{dv3h@virginia.edu}\affiliation{Department of Physics, University of Virginia,
Box 400714, Charlottesville, VA 22904, USA}
\date{November 3, 2016}
%
%

%
%

\begin{abstract}
A constraint of vanishing energy-momentum tensor  is motivated by a variety of perspectives on quantum gravity.  We  demonstrate in a concrete example how this constraint  leads to a metric-independent theory in which 
quantum gravity emerges as a nonperturbative artifact of regularization-scale physics. We analyze a scalar theory similar to the Dirac-Born-Infeld (DBI) theory with vanishing gauge fields, with the DBI Lagrangian modulated 
by a scalar potential. In the limit of a large number of scalars, we explicitly demonstrate the existence of a composite massless spin-2 graviton in the spectrum that couples to matter as in Einstein gravity. We comment on the 
cosmological constant problem and the generalization to theories with fermions and gauge fields. 
\end{abstract}
\maketitle

\section{Introduction}

The idea that gauge interactions and gravitation may arise as emergent phenomena has a long history. The earliest compelling model of emergent electromagnetism was due to Bjorken \cite{Bjorken:1963vg}, who proposed a 
scenario in which a four-fermion interaction of the form ${\cal L}_{\rm int}=G\,(\overline{\psi}\gamma^\mu\psi)(\overline{\psi}\gamma_\mu\psi)$ gives rise to a massless spin-1 composite state that interacts like the photon in electrodynamics. Bjorken 
argued that the dynamics of electromagnetism emerges in this scenario if the electromagnetic current, $J^\mu=\overline{\psi}\gamma^\mu\psi$, develops a nonvanishing condensate in the vacuum.
It was  further argued by Eguchi that Nambu-Jona-Lasinio-type models with emergent gauge interactions, such as Bjorken's model, may be renormalizable despite naive dimensional power counting as a consequence of 
the interactions between fundamental fermions and the collective gauge field excitations \cite{Eguchi:1976iz}. 

During the development of the theory of the strong interactions, when substantial evidence for color confinement as a theoretical consequence of quantum chromodynamics was still lacking, the possibility was briefly considered that color confinement  should be imposed via a constraint of vanishing color current on field configurations in an otherwise free theory of fermions  \cite{Amati:1974rm,Rajasekaran:1977td}. The full dynamics of quantum chromodynamics appeared to emerge as a consequence of confinement so implemented, rather than the other way around. 

Following the successful description of the electroweak and strong interactions by the Standard Model, a renormalizable gauge theory,  the existence of emergent gauge interactions was evidently unnecessary in order to explain existing experiments and observations. Still, it  remains a  possibility that some or all of the Standard Model gauge interactions are absent in a more fundamental description apparent only at short distances, yet  emerge in an effective description applicable to the low-energy environment in which  experiments have so far been performed \cite{Suzuki:2016aqj}.

In the context of gravity, for which there still does not exist a fundamental description known to be consistent with the Standard Model, the  paradigm that general relativity emerges as the effective description of a massless composite spin-2 state remains  compelling and has motivated a large number of investigations.  A review of some of the approaches to emergent gravity, including some of the ideas discussed here, was presented in Ref.~\cite{Sindoni:2011ej}.  Much of the activity in this area has been inspired by Sakharov's observation that the assumption of general covariance in a quantum field theory, including its regulator, is generally sufficient to produce a curved-space effective action which contains the Einstein-Hilbert term \cite{Sakharov:1967pk}. The Einstein-Hilbert action is induced by radiative corrections even if it is absent at tree level, and the gravitational coupling is related to the regularization scale. Numerous explicit calculations in  theories for which the spacetime metric is treated as an auxiliary field have demonstrated the validity of Sakharov's claim (see, for example, Refs.~\cite{Terazawa:1976xx,Akama:1977hr,Amati:1981rf,Amati:1981tu}).  However, in Sakharov's interpretation, Einstein's equations arise semiclassically, with the vacuum expectation value of the energy-momentum tensor acting as the source for an otherwise classical metric; the implications of Sakharov's observation for quantum gravity remain far from clear \cite{Visser:2002ew}. 

The observation that gauge interactions can emerge from a constraint of vanishing current  suggests by analogy that  gravitational interactions might emerge via a constraint of vanishing energy-momentum tensor.  This is the possibility that we put forward in this paper, though our implementation of the constraint does not precisely parallel the related proposal for quantum chromodynamics proposed in Ref.~\cite{Amati:1974rm}.  
Vanishing of the energy-momentum tensor is also motivated by a treatment of the spacetime metric as an auxiliary field, as in the Polyakov form of bosonic string theory \cite{Polyakov:1981rd}.
Furthermore, a constraint of vanishing energy and momentum is  in the spirit of the Wheeler-deWitt equation \cite{DeWitt:1967yk}, which promotes the vanishing of energy and momentum densities on states, constraints which follow from time- and space-reparametrization invariance, to the status of an axiom underlying quantum gravity. However, vanishing of the full energy-momentum tensor  is a stronger set of constraints than vanishing of the energy and momentum densities.  

The main purpose of this paper is to analyze a diffeomorphism-invariant theory of $N+D$ scalar fields which realizes the constraint of vanishing energy-momentum tensor and gives rise to an emergent gravitational interaction in $D$ spacetime dimensions.  
At the linearized level, the composite graviton couples to the energy-momentum tensor of $N$ of these scalars, while the remaining $D$ scalars are gauged away. A similar model was discussed  in Ref.~\cite{Akama:1978pg}.   Our work differs
from Ref.~\cite{Akama:1978pg}, not only in the initial assumptions that lead to the general form of the action, but also in that we directly demonstrate the existence of a graviton 
pole in a scattering amplitude at the nonperturbative level.   Avoiding the auxiliary field approach of Ref.~\cite{Akama:1978pg} allows us to address the issue of gauge fixing in a more transparent way and to directly identify the physical degrees of freedom that couple to the composite graviton in terms of the fields of the original theory.

Before proceeding, we should consider the feasibility of emergent gravitation in the first place. On the one hand, Weinberg demonstrated that a massless spin-2 state in a relativistic field theory must couple to matter as in Einstein gravity~\cite{Weinberg:1964ew}, 
so our task is to demonstrate the existence of such a state in the spectrum and check that it gives rise to gravitational interactions consistent with Einstein gravity.
However, Weinberg and Witten demonstrated the impossibility of massless  spin-2 states in  Lorentz-invariant field theories with an S-matrix and a nonvanishing conserved Lorentz-covariant energy-momentum 
tensor~\cite{Weinberg:1980kq,Porrati:2008rm}.  Any theory which claims to contain such massless spin-2 states must somehow violate the assumptions of the Weinberg-Witten theorem. The Weinberg-Witten theorem is proven by considering matrix elements of the conserved energy-momentum tensor between one-particle states of definite helicity. By identifying the effect of a spatial rotation of the energy-momentum tensor  on the one hand, and  rotation of the states on the other, one finds that the two rotations can be equivalent only if the massless states have helicity $\leq 1$. One possibility for evading the theorem is that the theory is diffeomorphism invariant from the outset, so that the energy-momentum tensor does not satisfy $\partial_\mu T^{\mu\nu}=0$. This is  how general relativity escapes the conclusions of the Weinberg-Witten theorem. The theory presented here is diffeomorphism invariant, but evades the Weinberg-Witten theorem even more directly: the energy-momentum tensor vanishes by construction. 

To demonstrate the emergent gravitational interaction, we analyze the  model at large $N$ and find a nonperturbative massless spin-2 pole in the Fourier transform of the $2\rightarrow 2$ scattering amplitude. While the classical theory is metric-independent and does not contain a nontrivial conserved energy-momentum tensor, after eliminating from functional integrals the integration over redundant degrees of freedom, we find that the composite graviton couples at the linearized level to the usual energy-momentum tensor of the surviving physical degrees of freedom. Up to regularization-scale-suppressed corrections, the effective low-energy theory contains scalar fields with  a potential, coupled to Einstein gravity. 

We present the theory in Section~\ref{sec:Theory}. We demonstrate the existence of a nonperturbative composite graviton for large $N$ in Section~\ref{sec:pole}. We discuss nonlinear matter-gravity couplings in Section~\ref{sec:nonlinear}. We comment on the cosmological constant problem (with more details in an appendix) and the possible generalization to a realistic theory of fermions and gauge interactions in Section~\ref{sec:Conclusions}.

\section{The Theory}
\label{sec:Theory}

We begin by considering the action for a collection of $N+D$ scalar fields in a curved $D$-dimensional spacetime described by a metric $g_{\mu\nu}$, where $\mu$, $\nu\in\{0,1,\dots,D-1\}$:
\begin{equation}
S=\int d^Dx\,\sqrt{|g|}\left[\frac{1}{2}g^{\mu\nu}\left(\sum_{a=1}^N\partial_\mu\phi^a\,\partial_\nu\phi^a+\sum_{I,J=0}^{D-1}\partial_\mu X^I\partial_\nu X^J\eta_{IJ}\right)
-V(\phi^a)\right], \label{eq:S1}
\end{equation}
where $g=\det(g_{\mu\nu})$, $g^{\mu\nu}$ is the inverse of the metric $g_{\mu\nu}$, and $\eta_{IJ}$ are constants which we take to have the values of the Minkowski metric in $D$ dimensions. (We use the mostly-minus convention for the signature of $\eta_{IJ}$ and the metric $g_{\mu\nu}$.)  The fields $X^1$, $X^2, \dots, X^{D-1}$  have ``wrong sign'' kinetic terms, but due to general coordinate invariance those fields may be gauge fixed  and are not independently dynamical, as we will discuss below. We assume   that the fields $X^I$ do not appear in the potential $V(\phi^a)$ so that the action maintains a global shift symmetry $X^I(x)\rightarrow X^I(x)+\Delta^I$.  The action (\ref{eq:S1}) is also reparametrization invariant with the fields $X^I, \phi^a$ transforming as scalars and the background field $g_{\mu\nu}$ transforming as a metric tensor. We want to emphasize at this point that classically the background metric $g_{\mu\nu}$ has no dynamics.

We proceed to define the theory by  functional integral quantization over the fields $\phi^a(x)$, $X^I(x)$ and $g_{\mu\nu}(x)$, {\em subject to the constraint that the energy-momentum tensor vanishes:} $T_{\mu\nu}(x)=0$.  The partition function for the theory is \begin{equation}
Z=\int_{T_{\mu\nu}=0} {\cal D}g_{\mu\nu}\, {\cal D}\phi^a\, {\cal D}X^I\,e^{iS[\phi^a,X^I,g_{\mu\nu}]}.
\end{equation}
The energy-momentum tensor is given by \begin{eqnarray}
T_{\mu\nu}(x)&=&\frac{2}{\sqrt{|g|}}\frac{\delta S}{\delta g^{\mu\nu}(x)} \label{eq:Tmn} \\
&=&\sum_{a=1}^N\partial_\mu \phi^a \partial_\nu\phi^a +\sum_{I,J=0}^{D-1}\partial_\mu X^I\partial_\nu X^J\eta_{IJ}-g_{\mu\nu} {\cal L},\label{eq:Tmn2}\end{eqnarray}
where the Lagrangian ${\cal L}$ is defined by the action in Eq.~(\ref{eq:S1}), $S\equiv\int d^Dx\,\sqrt{|g|}{\cal L}$.

The constraint $T_{\mu\nu}=0$ determines the spacetime metric in terms of the scalar field configuration. Each field configuration contributes to the functional integral as if it propagates in a unique spacetime unrelated to Einstein's equations.
An explicit solution to $T_{\mu\nu}(x)=0$ for the metric is
\begin{equation}
g_{\mu\nu}=\frac{D/2-1}{V(\phi^a)}\left(\sum_{a=1}^N\partial_\mu \phi^a\partial_\nu\phi^a+
\sum_{I,J=0}^{D-1}\partial_\mu X^I\partial_\nu X^J\eta_{IJ}\right). 
\label{eq:gmn}\end{equation}
For generic field configurations and potentials, $g_{\mu\nu}$ in Eq.~(\ref{eq:gmn}) is nonsingular.  In the perturbative expansion about a Minkowski-space background that we employ later, difficulties
associated with singularities in the metric do not appear\footnote{A similar issue regarding the singular nature of the induced metric appears in string theory. Though most of the time the D-brane action is taked to be non-singular, Gibbons and Ishibashi \cite{Gibbons:2004dz} have shown that there exist classical D-brane configurations with vanishing gauge fields (specifically D3 branes embedded in a 5-dimensional Lorentzian spacetime) whose induced metric exhibits signature change.  The authors further noted that despite the curvature singularities exhibited by the induced metric, from the point of the bulk the D-brane geometry is smooth everywhere. Here we will ignore such singular solutions. }.

With the spacetime metric determined by Eq.~(\ref{eq:gmn}), the action for the theory resembles the  Dirac-Born-Infeld action with vanishing gauge field, modulated by the scalar-field potential function $V(\phi^a)$:
\begin{equation}
S=\int d^Dx\ \left(\frac{\tfrac D2-1}{V(\phi^a)} \right)^{\frac{D}{2}-1}
\sqrt {\bigg|\det \left(\sum_{a=1}^N \partial_\mu\phi^a \,\partial_\nu\phi^a 
+\sum_{I,J=0}^{D-1}\partial_\mu X^I \,\partial_\nu X^J\, \eta_{IJ}\right)\bigg|}.
\label{eq:S}\end{equation}
The action is now independent of the spacetime metric, by construction. As a consequence, the energy-momentum tensor defined as in Eq.~(\ref{eq:Tmn}) vanishes identically. Equivalently, one can confirm that the Noether currents associated with space- and time-translation invariance vanish.

The functional integrals that determine the partition function and correlation functions in this theory include the redundant integration over field configurations related by coordinate transformations. In order to compute correlation functions in the theory, we gauge fix the spacetime parametrization by identifying the fields $X^I$ with the corresponding spacetime coordinates, in analogy with the static gauge condition in string theory, up to an overall constant factor:
\begin{equation}
X^I= \sqrt{\frac{V_0}{(\tfrac{D}{2}-1)}} \ x^\mu\delta_\mu^I , \ \ I=0,\dots,D-1. \label{eq:staticgauge}
\end{equation}
Here, $V_0$ is a dimensionful parameter that will be chosen later. Note that this gauge condition would not be possible
if there were fewer than $D$ scalar fields.
In the functional integral we can include the identity in the form,
\begin{equation}
1=\int {\cal D}\alpha^\mu(x) \,\delta\left(X^{I,\alpha}(x)- \sqrt{\frac{V_0}{(\tfrac{D}{2}-1)}} x^I\right)\,\det\left(\frac{\delta X^{I,\alpha}(y)}
{\delta \alpha^\mu(y')}\right) ,
\label{eq:1}\end{equation}
where  $X^{I,\alpha}(x^\mu)\equiv X^I(x^\mu+\alpha^\mu(x))$.
Note that the Fadeev-Popov determinant is trivial here, and there are no ghosts associated with the static gauge condition:
\begin{equation}
\frac{\delta X^{I,\alpha}(y)}{\delta \alpha^\mu(y')}=\partial_\mu X^I(y)\,\delta^{(D)}(y-y')= \sqrt{\frac{V_0}{(\tfrac{D}{2}-1)}}  \delta^I_\mu\,\delta^{(D)}(y-y'),  \label{eq:FP}\end{equation}
where the last identity in Eq.~(\ref{eq:FP}) is a consequence of the gauge-fixing condition. The integrals over coordinate-transformation functions 
$\alpha^\mu(x)$ then factor out of correlation functions of coordinate-invariant observables.
Locally we can always transform to spacetime coordinates $x^\mu$ which satisfy the static gauge condition (\ref{eq:staticgauge}).
However, globally this choice imposes a topological constraint on the space of allowed field configurations, a constraint which we will impose in our perturbative analysis, but which should not be necessary for the emergence of gravitation in the theory.   We also note that in static gauge the field $X^0$ inherits the role of an internal clock \cite{Page:1983uc}, and the fields $X^i$, $i=1,\dots D-1$ rulers. The clock and ruler fields provide physical meaning to the spacetime backdrop in which dynamics takes place.   It is sensible to consider correlation functions of operators built out of the local fields $\phi^a(x^\mu)$ after gauge fixing, even though those correlation functions take a different functional form in other gauge choices\footnote{This is analogous to saying that it is sensible to consider particle trajectories in general relativity, even though 
the trajectory is described differently in different coordinate systems.}.

The classical equations of motion for the fields $\phi^a$ and $X^I$ following from the action Eq.~(\ref{eq:S}) may be written,
\begin{eqnarray}
\frac{1}{\sqrt{|g|}}\partial_\mu\left(\sqrt{|g|}g^{\mu\nu}\partial_\nu\phi^a\right)=-\frac{\partial V}{\partial \phi^a}, \label{eq:phiEOM} \\
\frac{1}{\sqrt{|g|}}\partial_\mu\left(\sqrt{|g|}g^{\mu\nu}\partial_\nu X^I\right)=0 , \label{eq:XEOM}
\end{eqnarray}
where the composite operator $g_{\mu\nu}$ that  plays the role of the spacetime metric is precisely the solution to $T_{\mu\nu}=0$ in Eq.~(\ref{eq:gmn}).
As a result of the assumed shift symmetry which acts on the fields $X^I$, the static gauge choice Eq.~(\ref{eq:staticgauge}) satisfies the  equations of motion, Eqs.~(\ref{eq:phiEOM}) and (\ref{eq:XEOM}), with  fields $\phi^a$ uniform and fixed to the minimum of the potential $V(\phi^a)$, and with $g_{\mu\nu}$ a constant tensor proportional to $\eta_{\mu\nu}$. This suggests that an  expansion in the  fields $\phi^a$ may remain perturbative in this gauge choice. We will find this to be the case, up to the expected instability of the Minkowski-space solution if a regularization-scale-dependent cosmological constant is not tuned to zero. 
  
\section{Graviton Pole} \label{sec:pole}

In order to analyze the theory perturbatively,  we write $V(\phi)=V_0+\Delta V(\phi^a)$ and expand the action Eq.~(\ref{eq:S}) in powers of $1/V_0$. We also assume that $N$, the number of fields $\phi^a$ in the theory, is large, and keep only leading terms in a $1/N$ expansion. For the moment we identify the constant $V_0$ in the potential with the $V_0$ that appeared in the gauge-fixing condition Eq.~(\ref{eq:staticgauge}). We modify this condition later.

We write the gauge-fixed action as follows:
\begin{equation}
S=\int d^Dx\,\frac{V_0}{D/2-1}\left(\frac{V_0}{V_0+\Delta V(\phi^a)}\right)^{D/2-1}\sqrt{\bigg|\det\left(\eta_{\mu\nu}+\tilde{h}_{\mu\nu}\right)\bigg|}, \label{eq:s1}\end{equation}
where 
\begin{equation}
\tilde{h}_{\mu\nu}\equiv \frac{\tfrac D2-1}{V_0}\left( \sum_{a=1}^N\partial_\mu\phi^a\partial_\nu\phi^a \right), \label{eq:htilde}\end{equation}
and \begin{equation}
g_{\mu\nu}=\frac{V_0}{V(\phi)}\left(\eta_{\mu\nu}+\tilde{h}_{\mu\nu}\right)
\end{equation}
in the static gauge Eq.~(\ref{eq:staticgauge}).  Expanding the determinant using $\det M =\exp({\rm Tr}\,{\rm ln}\,M)$, we have the expansion 
\begin{equation}
S=\int d^Dx\,\frac{V_0}{D/2-1}\,\left(1+\frac{\Delta V(\phi^a)}{V_0}\right)^{1-D/2}\left(1+\frac{1}{2}\tilde{h}_\mu^\mu-
\frac{1}{4}\tilde{h}_{\mu\nu}\tilde{h}^{\mu\nu}+\frac{1}{8}(\tilde{h}_\mu^\mu)^2+\cdots\right),\end{equation}
where index contractions are done with the Minkowski metric $\eta_{\mu\nu}$. Expanding  the factor involving $\Delta V/V_0$, and using Eq.~(\ref{eq:htilde}), we find that the action can be written as
\begin{eqnarray}
S=\int d^Dx&&\left\{\frac{V_0}{D/2-1}+\frac{1}{2}\sum_{a=1}^N \partial_\mu \phi^a \partial^\mu \phi^a -\Delta V(\phi^a )\right. \nonumber \\ &&
-\frac{\tfrac D2-1}{4V_0}\left[
\sum_{a=1}^N\partial_\mu\phi^a \partial_\nu\phi^a 
\,\sum_{b=1}^N\partial^\mu\phi^b\partial^\nu \phi^b
-\frac{1}{2}\left(\sum_{a=1}^N \partial_\mu \phi^a \partial^\mu \phi^a  \right)^{\!\!2\,}\right] \nonumber \\
&& \left.-\frac{\tfrac{D}{2}-1}{2}\frac{\Delta V(\phi^a)}{V_0}\sum_{a=1}^N \partial_\mu \phi^a\partial^\mu \phi^a+\frac{D}{4}\frac{(\Delta V(\phi^a))^2}{V_0}+{\cal O}\left(\frac{1}{V_0^2}\right)\right\}.   \label{eq:Sexpansion}
\end{eqnarray}
For now, we assume a free theory with O$(N)$-symmetric potential 
\begin{equation}
\Delta V(\phi^a)=\sum_{a=1}^N\frac{m^2}{2}\phi^a\phi^a,\end{equation}
in which case the first line of Eq.~(\ref{eq:Sexpansion}) is the free part of the action.
In terms of the energy-momentum tensor for free fields $\phi^a$ in Minkowski space\footnote{Note that the energy-momentum tensor that will act as the source for linearized gravity, ${\cal T}_{\mu\nu}$, is not the full energy-momentum tensor of the theory, $T_{\mu\nu}$ in Eq.~(\ref{eq:Tmn2}), which vanishes.},
\begin{equation}
{\cal T}^{\mu\nu}= \sum_{a=1}^N\bigg[\partial^\mu \phi^a \partial^\nu \phi^a - \eta^{\mu\nu} \left(\frac{1}{2} \partial^\alpha \phi^a \partial_\alpha \phi^a - \frac{1}{2} m^2 \phi^a \phi^a \right)\bigg]  \,\,\, ,
\label{eq:tflat}
\end{equation}
 the interacting terms to ${\cal O}(1/V_0)$ in the Lagrangian take the simple form, 
 \begin{equation}
{\cal L}_{int} =-\frac{1}{4 V_0} {\cal T}_{\mu\nu} {\cal T}_{\alpha\beta} \left[ (D/2-1) \, \eta^{\nu\alpha} \eta^{\mu\beta} - \frac{1}{2} \eta^{\mu\nu} \eta^{\alpha\beta} \right] \,\,\, .
\label{eq:lint}
\end{equation}
 
We will now demonstrate that this theory includes a massless spin-two graviton state that mediates a gravitational interaction between $\phi^a$ particles. 
We consider two-into-two scattering of $\phi$ bosons in the large $N$ limit to determine first whether the scattering amplitude has a massless pole; we then study the tensor structure of the 
associated propagator to show that it has the  form appropriate for Einstein gravity.

\begin{figure}[t]
  \begin{center}
    \includegraphics[width=.65\textwidth]{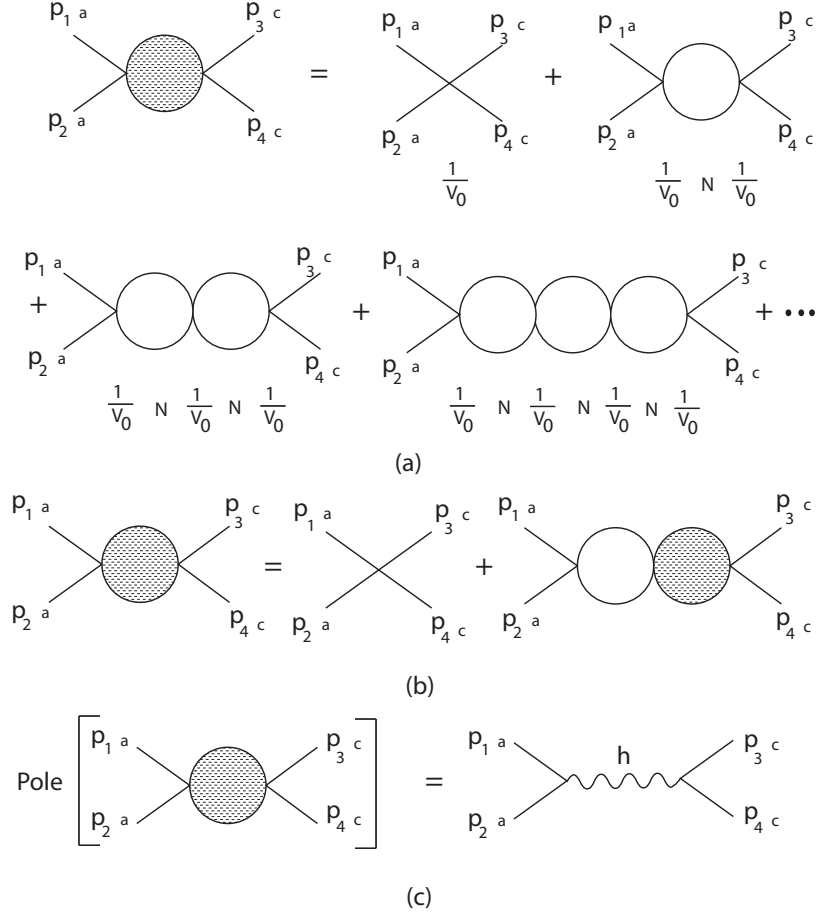}
    \caption{Two-into-two scattering of $\phi$ bosons. (a) Diagrams that contribute at leading order in $1/N$, for the choice of external fields
    described in the text. (b) The equivalent recursive representation. (c) The effective graviton exchange diagram.}
        \label{fig:scat}
  \end{center}
\end{figure}
The scattering amplitude is represented by ${\cal M}(p_1,a \,; p_2,b \rightarrow p_3,c\,; p_4,d)$, where $a$, $b$, $c$ and $d$ indicate which of the $N$ scalar fields participate
in the scattering process.  We make the choice $a=b \neq c=d$;  the relevant Feynman diagrams that contribute to this amplitude at leading order in $1/N$ are shown in Fig.~\ref{fig:scat}a.
For the given choice of external lines, $t$- and $u$-channel diagrams are subleading in $1/N$. The order $N$ enhancement present for the $s$-channel diagram comes from tracing over the flavor degrees of the scalars running in the loop.  

The reader may worry at this point that we have omitted many other possible diagrams that contribute at the same order in $1/N$; for example, one
could take any diagram shown in Fig.~\ref{fig:scat}a and simply append a single loop to any one of the internal lines.  We will  show that {\em all}
other diagrams that might contribute to the scattering amplitude at leading order in $1/N$ are cancelled as a consequence of the way in which we impose the condition of vanishing cosmological constant.  This constraint is necessary for the stability of the theory given that we have chosen a
Minkowski-space background.  We discuss the case where the cosmological constant is non-vanishing in an appendix.

Let us parameterize the two degrees of freedom in our theory that are relevant to this discussion.  First, the background value of the fields $X^I$ need not have been chosen as in
Eq.~(\ref{eq:staticgauge});  we now allow for a rescaling of $X^I$,
\begin{equation}
X^I=  x^I\, \sqrt{\frac{V_0}{(\tfrac{D}{2}-1)}-c_1}, \   \label{eq:staticgauge2}
\end{equation}
where $c_1$ is a constant. This change modifies the form of $\tilde{h}_{\mu\nu}$ in Eq.~(\ref{eq:htilde}) so that the operator $\partial_\mu \phi^a \partial_\nu \phi^a$ is replaced by  
$\partial_\mu \phi^a \partial_\nu \phi^a - c_1 \eta_{\mu\nu}$.   In addition, while still expanding the action in powers of $1/V_0$, we 
add an additional constant to the scalar potential, so that the original 
$\Delta V$ is replaced by 
\begin{equation}
\Delta V = m^2 \phi^a \phi^a/2 - c_2 \,\,\, .   
\label{eq:dvnew}
\end{equation}
Since the action of our theory is exclusively a function 
of $\tilde{h}_{\mu\nu}$ and $\Delta V$, these modifications assure that every occurrence of the operator $\partial_\mu \phi^a \partial_\nu \phi^a$ is via 
$\partial_\mu \phi^a \partial_\nu \phi^a - c_1 \eta_{\mu\nu}$ and every occurrence of $m^2 \phi^a \phi^a/2$ is via $m^2 \phi^a \phi^a/2 - c_2$.   We may think
of $c_1$ and $c_2$ as counterterms: any diagram involving a loop that is created by connecting a $\phi^a$ line within the operator $\phi^a \phi^a$ or within
$\partial_\mu \phi^a \partial_\nu \phi^a$ will be paired with another diagram that has a counterterm vertex in place of that loop.  We impose the renormalization
condition that the sum of these diagrams vanishes.  This is merely a convenient way to organize the perturbative calculation, since a different choice of
the counterterms is equivalent to a change in gauge and a shift in $V_0$, which we have not yet fixed. In our scattering problem, this eliminates all other diagrams 
that are leading order in $1/N$ that are not shown in Fig.~\ref{fig:scat}a.  Our renormalization condition is equivalent to replacing the operators $\phi^a\phi^a$ and 
$\partial_\mu \phi^a \partial_\nu \phi^a$ in Eqs.~(\ref{eq:tflat}) and (\ref{eq:lint}) with normal-ordered operators $: \!\phi^a \phi^a \!\!:$ and $:\!\partial_\mu \phi^a \partial_\nu \phi^a\!\!:$, respectively.

What is convenient about the problem as we have set it up is that the diagrams in Fig.~\ref{fig:scat}a can be represented by the recursive formula shown graphically in Fig.~\ref{fig:scat}b.\footnote{A similar analysis in a theory with emergent electromagnetic interactions was performed in Ref.~\cite{Suzuki:2016aqj}.}
Each diagram in the infinite sum involves common functions of the external momenta that we will factor out in doing the resummation.  We define
\begin{equation}
E_{\mu\nu}(p_1,p_2) \equiv - (p_1^\mu \, p_2^\nu + p_1^\nu \, p_2^\mu) + \eta^{\mu\nu} (p_1 \cdot p_2 +m^2) \,\,\, ,
\label{eq:exln}
\end{equation}
for inwardly (or outwardly) directed external momenta $p_1$ and $p_2$.   At leading order in $1/N$, the Feynman rules for the external lines follow solely from the action of ${\cal T}^{\mu\nu}$
on the external states; hence the form of Eq.~(\ref{eq:exln}) is determined by Eq.~(\ref{eq:tflat}), summing over the ways in which the fields can annihilate (or create) incoming (or outgoing) scalar 
bosons.  We then write the scattering amplitude in the form  
\be
i {\cal M} (p_1,a \,; p_2,a \rightarrow p_3,c\,; p_4,c)\equiv E_{\mu\nu}(p_1,p_2) [i \,A^{\mu\nu|\rho\sigma}(q) ] \,E_{\rho\sigma}(p_3,p_4)\,\,\, ,
\ee
where 
\be
q=p_1+p_2=p_3+p_4.
\ee  
The recursive relation represented by Fig.~\ref{fig:scat}b is given by
\begin{equation}
A^{\mu\nu|\rho\sigma}(q) = A^{\mu\nu|\rho\sigma}_0 + {K^{\mu\nu}}_{\alpha\beta}(q)\, A^{\alpha\beta|\rho\sigma}(q)  \,\,\,.
\label{eq:recursive}
\end{equation}
The first term on the right-hand-side follows from the tree-level amplitude and is given by
\begin{equation}
A^{\mu\nu|\rho\sigma}_0 = - \frac{1}{4 V_0} \left[ (\tfrac D2-1) \, \left(\eta^{\nu\rho} \eta^{\mu\sigma}+ \eta^{\mu\rho} \eta^{\nu\sigma}\right) - \eta^{\mu\nu} \eta^{\rho\sigma} \right]  \,\,\, ,
\label{eq:tamp}
\end{equation}
while the kernel ${K^{\lambda\kappa}}_{\rho\sigma}(q)$ can be found by doing a one-loop calculation corresponding to the portion of the second diagram in Fig.~\ref{fig:scat}b that 
connects to the shaded blob.  We find
\begin{equation}
{K^{\lambda\kappa}}_{\rho\sigma} = -\frac{i N}{4 V_0} \left[ (\tfrac{D}{2}-1) \, \eta^{\kappa\alpha} \eta^{\lambda\beta} - \frac{1}{2} \eta^{\lambda\kappa} \eta^{\alpha\beta} \right]
\int \frac{d^D l}{(2 \pi)^D} \frac{E_{\alpha\beta}(\frac{q}{2}-l,\frac{q}{2}+l) \, E_{\rho\sigma}(\frac{q}{2}-l,\frac{q}{2}+l)}{[(\frac{q}{2}-l)^2-m^2][(\frac{q}{2}+l)^2-m^2]}  \,\,\, . \label{kern1}
\end{equation}
In the appendices, we present the calculation of the kernel.   The result may be written as Eq.~(\ref{kernel}),
\begin{equation}
{K^{\lambda\kappa}}_{\rho\sigma} = \frac{N(D/2-1)}{4 V_0} \frac{\Gamma(-D/2)}{(4\pi)^{D/2}} (m^2)^{D/2} \left[1-\frac{D}{12} \frac{q^2}{m^2}\right]\left(\delta^\kappa_\rho \, 
\delta^\lambda_\sigma + \delta^\kappa_\sigma \,
\delta^\lambda_\rho\right) + {\cal O}(q^4)\,\,\, ,
\label{eq:kernalsanders}
\end{equation}
where we have omitted terms that vanish as a consequence of the identity\footnote{It is also true that $q_\mu \, E^{\mu\nu}(p_3,p_4) = 0$.}
\begin{equation}
q_\mu \, E^{\mu\nu}(p_1,p_2) = 0  \,\,\, .\label{transv}
\end{equation}
Defining the quantity $\lambda$ by
\begin{equation}
\lambda \equiv \frac{N(D/2-1)}{2 V_0} \frac{\Gamma(-D/2)}{(4\pi)^{D/2}} (m^2)^{D/2} \,\,\, ,
\end{equation}
we may substitute our solution for the kernel into Eq.~(\ref{eq:recursive}), from which we obtain
\begin{equation}
 (1-\lambda) A^{\mu\nu|\rho\sigma}(q) = A^{\mu\nu|\rho\sigma}_0 + \lambda \left(\frac{D}{12} \frac{q^2}{m^2} \right)A^{\mu\nu|\rho\sigma}(q) + {\cal O}(q^4) \,\,\, .
 \label{eq:subin}
\end{equation}
We wish to determine whether the existence of a massless pole can be inferred from Eq.~(\ref{eq:subin}).  We make the fine-tuned choice \begin{equation}
V_0=\frac{N(D/2-1)}{2} \frac{\Gamma(-D/2)}{(4\pi)^{D/2}} (m^2)^{D/2} \,\,\, ,
\label{eq:special}
\end{equation}
which corresponds to $\lambda=1$.\footnote{With this choice of $V_0$, we find at one loop that $c_1=-V_0/(\tfrac D2-1)$, so that in  Eq.~(\ref{eq:staticgauge2}) the fields $X^I$ remain real, with
$D=4-\epsilon$ and $\epsilon>0$.}  In this case, the left-hand-side of Eq.~(\ref{eq:subin}) vanishes, and it immediately follows that the amplitude contains a massless pole:
\begin{equation}
A^{\mu\nu|\rho\sigma}(q) =  \frac{3 \,m^2}{D\, V_0}\, \left[(\tfrac D2-1) \,( \eta^{\nu\rho} \eta^{\mu\sigma} + \eta^{\nu\sigma} \eta^{\mu\rho}) 
- \eta^{\mu\nu} \eta^{\rho\sigma} \right] \, \frac{1}{q^2} +\cdots \,\,\, ,
\label{eq:ampsol}
\end{equation}
where we have used the tree-level amplitude given by Eq.~(\ref{eq:tamp}).  Notice that Eq.~(\ref{eq:ampsol}) has the tensor structure that one expects for the graviton propagator of Einstein gravity in de Donder gauge.

Our choices of $c_1$, $c_2$ and $V_0$ are tuned to give a vanishing cosmological constant.  Notice that if we were to hold $c_1$ and $c_2$ fixed while 
replacing $V_0 \rightarrow V_0 + \delta V_0$, we would find that $\delta V_0$ appears in two ways in the action: (1) in a constant that multiplies every 
occurrence of the Minkowski metric, which can be removed by a rescaling of the fields and of $m^2$ and (2) in a shift of the scalar potential.  In 
Appendix~\ref{App:CC}, we show that the latter indicates the presence of a cosmological constant and we discuss the consequences for the theory.​

We now return to the physical interpretation of our renormalization conditions on the counterterms $c_1$ and $c_2$.  Consider first how gravity couples to matter in our theory.   Using the solution for the metric $g_{\mu\nu}$, we may identify the composite operator that represents the fluctuation about Minkowski space; it can be expressed in terms of ${\cal T}^{\mu\nu}$ as 
\begin{equation}
h_{\lambda\kappa} = \frac{1}{V_0} {P^{\alpha\beta}}_{\lambda\kappa} \, {\cal T}_{\alpha\beta} + {\cal O}(1/V_0^2) \,\,\, ,
\label{eq:HtoT}
\end{equation}
where we define the tensor structure
\begin{equation}
 {P^{\alpha\beta}}_{\lambda\kappa} \equiv \frac{1}{2}  \left[ (\tfrac D2-1)\left(\delta^\alpha_\lambda \delta^\beta_\kappa + \delta^\alpha_\kappa \delta^\beta_\lambda\right)
 - \eta_{\lambda\kappa}\eta^{\alpha\beta} \right] \,\,\, . \label{eq:P}
 \end{equation}
To determine the couplings to $\phi$ that are linear in  $h_{\alpha\beta}$ in an effective theory in which $h_{\alpha\beta}$ is treated as a 
fundamental field, we note that there are two possible identifications of the composite operator $h_{\mu\nu}$ in Eq.~(\ref{eq:lint}) that are relevant.   Including 
both, we invert Eq.~(\ref{eq:HtoT}) and substitute, yielding
\begin{equation}
{\cal L}_{eff} \supset -\frac{1}{2} h^{\alpha\beta} {\cal T}_{\alpha\beta}  \,\,\, ,
\label{eq:ourex}
\end{equation}
which matches our expectation for the graviton coupling implied by linearized general relativity.  At the order in $1/V_0$ that we are working, a cosmological constant would be present if there were a non-vanishing 
vacuum expectation value proportional to $\eta_{\alpha\beta}$ for the operator ${\cal T}_{\alpha\beta}$.   However,  we have imposed renormalization conditions on $c_1$ and $c_2$ so that diagrams formed by closing a $\phi^a$ line in either the 
operator $\phi^a\phi^a$ or $\partial_\alpha \phi^a \partial_\beta \phi^a$ vanish identically.    The diagrams that would contribute to $\langle {\cal T}_{\alpha\beta} \rangle$ are of this form;  hence our renormalization 
choice fixes the cosmological constant to zero, to the order in $1/V_0$ that we are working.   In Appendix~\ref{App:CC} we consider the consequences of not imposing this tuning and we compare 
to the expectations of general relativity when one tries to expand about a Minkowski-space background in the presence of a cosmological constant.

Finally, we may use our results to compute the $D$-dimensional Planck mass $M_{\Pl}$. The corresponding scattering amplitude 
computed via linearized general relativity is given by\footnote{In general relativity the scattering amplitude follows from the linearized gauge-fixed action
$${\cal S}=\int d^D x \bigg[-\tfrac 12 h_{\mu\nu}\bigg(\tfrac 12 (\eta^{\mu\alpha}\eta^{\nu\beta}+\eta^{\mu\beta} \eta^{\nu\alpha})-\tfrac 12 \eta^{\mu\nu}\eta^{\alpha\beta}\bigg)\Box h_{\alpha\beta}+M_{\Pl}^{1-D/2}h_{\mu\nu} {\cal T}^{\mu\nu}\bigg]\,\,\, ,$$ where we parametrized the metric fluctuation around flat space by $\delta g_{\mu\nu}=2M_{\Pl}^{1-D/2} h_{\mu\nu}$ and introduced the source coupling via the standard definition ${\cal T}^{\mu\nu}=(2/\sqrt{|g|}) \delta {\cal L}/\delta g_{\mu\nu}$. This action is obtained by linearizing the Einstein-Hilbert term $\int d^D x (\tfrac 12 \partial_\rho h_{\mu\nu} \partial^\rho h^{\mu\nu}-\partial_\mu h_{\nu\rho}\partial^\nu h^{\mu\rho}+\partial_\mu h^{\mu\nu}\partial_\nu h - \tfrac 12 \partial_\mu h \partial^\mu h)$ and adding to it the de Donder gauge-fixing term $\int d^D x (\partial^\nu h_{\mu\nu}-\tfrac 12 \partial_\mu h)^2,$ and the coupling to matter. The propagator in the de Donder gauge is $i/(D/2-1)\, P_{\mu\nu|\alpha\beta}/q^2,$ where $P_{\mu\nu|\alpha\beta}$ was previously defined in Eq.~(\ref{eq:P}). } 
\begin{equation}
A^{\mu\nu|\rho\sigma}(q) = \frac{M_{\Pl}^{2-D}}{D-2} \, \left[(\tfrac D2-1) \,( \eta^{\nu\rho} \eta^{\mu\sigma} + \eta^{\nu\sigma} \eta^{\mu\rho}) 
- \eta^{\mu\nu} \eta^{\rho\sigma} \right] \, \frac{1}{q^2} \,\,\, ,
\end{equation}
Comparing with Eq.~(\ref{eq:ampsol}), we identify
\begin{equation}
M_{\Pl}= m\,\bigg[\frac{- N \, \Gamma(1-\frac{D}{2})}{6\, (4 \pi)^{D/2} }\bigg]^{1/(D-2)} \,\,\, .
\end{equation}
It is important to keep in mind that we use dimensional regularization here as a placeholder for a generally covariant, physical regulator, associated with some high scale 
$\Lambda$.   We do so by writing $D=4-\epsilon$, with $\epsilon$ small, but finite and positive.   In this case, the four-dimensional Planck mass  is proportional to  $\sqrt{N/\epsilon} \,m $, which 
can be made large even when $m$ is comparable to the masses of known particles. We comment on other physical regulators in the Discussion and Conclusions of this paper.

\section{Graviton couplings beyond linear order}
\label{sec:nonlinear}

In this section, we begin addressing the question of how the non-linear graviton couplings in our model match one's expectations from
general relativity.   We focus primarily on the couplings that originate from the matter action; a treatment of the graviton self-interactions,
{\em i.e.}, those originating from $\sqrt{|g|} \,R$ in general relativity, will be discussed below briefly, but studied in a longer 
publication~\cite{nextpaper}.

Prior to the imposition of the condition that the energy-momentum tensor vanishes, our initial action is that of a real scalar field in
curved space.   Expanding about the Minkowski-space background, $g_{\mu\nu} = \eta_{\mu\nu} + h_{\mu\nu}$,  we find the 
usual $h_{\mu\nu}$ couplings to $\phi^a$ that one would expect in general relativity.   In our case, however, the subsequent 
constraint we impose on the energy-momentum tensor identifies $h_{\mu\nu}$ with a function of the scalar fields $\phi^a$.  Using a diagrammatic argument below, we argue that it is justified to identify this composite operator as the operator that creates and annihilates gravitons;
then the theory includes all the multi-graviton couplings that we expect to find in $\sqrt{|g|} \,{\cal L}_{matter}$.
The theory contains more than just gravitational interactions; however, we expect the higher-dimensional local effective interactions generated by integrating over scalar loops to be suppressed by the regularization scale. An analysis of these operators will be presented in Ref.~\cite{nextpaper}.

\begin{figure}[t]
  \begin{center}
    \includegraphics[width=0.65\textwidth]{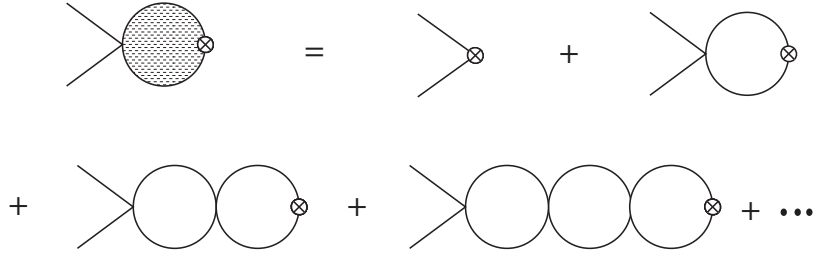}
    \caption{Chain of scalar loop diagrams connecting to the composite operator 
       $h_{\mu\nu}$. The operator appears in an interaction vertex that is
    represented by the small circled cross.}
        \label{fig:interp}
  \end{center}
\end{figure}

Consider any term in the expansion of $\sqrt{|g|} \,{\cal L}_{matter}$ written in terms of the $\phi^a$ and $h_{\mu\nu}$, prior to imposing our 
condition that the energy-momentum tensor vanishes.  We wish to focus on one factor of $h_{\mu\nu}$ appearing in such a term at a time, so 
we will represent the generic interaction as
\begin{equation}
{\cal L}_{matter} \supset -\frac{1}{2} h_{\mu\nu} \, {\cal V}^{\mu\nu}  \,\,\, ,
\label{eq:vh}
\end{equation}
where the factor of $-1/2$ is included as a matter of convenience and ${\cal V}^{\mu\nu} \equiv {\cal V}^{\mu\nu}(\phi, h_{\alpha\beta})$, with 
the chosen $h_{\mu\nu}$ factored out.  After we constrain the energy-momentum tensor of the original theory,  we may express $h_{\mu\nu}$ in 
terms of the flat-space energy-momentum tensor, as given by Eq.~(\ref{eq:HtoT}).  Hence, we can write Eq.~(\ref{eq:vh}) in the form
\begin{equation}
{\cal L}_{matter} \supset 2\, {\cal T}^{\mu\nu} \left( -\frac{1}{4V_0}  P_{\mu\nu|\alpha\beta} \right) {\cal V}^{\alpha\beta} + {\cal O}(1/V_0^2) \,\,\, ,
\label{eq:vpt}
\end{equation}
which we should compare to the form of the quartic interactions in Eq.~(\ref{eq:lint}),
\begin{equation}
{\cal L}_{int}  = {\cal T}^{\mu\nu} \left(-\frac{1}{4V_0}  P_{\mu\nu|\alpha\beta} \right) {\cal T}^{\alpha\beta} + {\cal O}(1/V_0^2) \,\,\, .
\label{eq:tpt}
\end{equation}
Now we consider the effect of  attaching a chain of scalar loops to the ${\cal T}^{\alpha\beta}$ that we obtained by
substituting for the designated $h_{\mu\nu}$.   The relevant diagrams are shown Fig.~\ref{fig:interp}.   The circled cross represents the Feynman rule 
following from $ {\cal V}^{\mu\nu}(\phi, h_{\alpha\beta})$, which we will represent by the function $E^V(p_3,\ldots, p_n)^{\mu\nu}$, assuming there
are $n$ external lines in the diagram (with $n-2$ not displayed in Fig.~\ref{fig:interp}).  

Since each diagram will involve a factor of $E_{\rho\sigma}$ on the left and $E^V_{\mu\nu}$on the right, we notice that the resulting amplitude has the form
 \begin{equation}
 i {\cal M} \equiv E_{\mu\nu}(p_1,p_2) [ i A^{\mu\nu | \rho\sigma}(q)] E^V_{\rho\sigma}(p_3,\ldots, p_n) \,\,\, ,
 \label{eq:sumofdiag}
 \end{equation}
 where $A^{\mu\nu|\rho\sigma}$ is the {\em same} amplitude defined in our previous two-into-two scattering calculation.  Note that the factor of $E_{\mu\nu}$ 
 on the left is sufficient to eliminate the same terms dropped from the kernel in Eq.~(\ref{eq:kernalsanders}).  In comparing
 Fig.~\ref{fig:scat} and Fig.~\ref{fig:interp},  the difference between Eqs.~(\ref{eq:vpt}) and (\ref{eq:tpt}) simply changes how each diagram
 in these figures terminates on the right.  The part of $A^{\mu\nu|\rho\sigma}$ that has a pole, Eq.~(\ref{eq:ampsol}), was shown in the previous section 
 to correspond to the graviton propagator.  Hence, if one is interested only in the graviton coupling, the result in Eq.~(\ref{eq:sumofdiag}) matches 
 what one would find by treating the designated $h_{\mu\nu}$ in Eq.~(\ref{eq:vh}) as a fundamental field.  Note that the factor of $E_{\rho\sigma}$ 
 associated with the left-side of each diagram is what we expect given the form of Eq.~(\ref{eq:ourex}).  Since the $h_{\mu\nu}$ appear 
 in $\sqrt{|g|} \,{\cal L}_{matter}$ exactly as in general relativity, the effective interactions for the composite graviton match with those that we would 
 obtain in the conventional formulation where $h_{\mu\nu}$ is identified as a fundamental field.

\begin{figure}[t]
  \begin{center}
    \includegraphics[width=0.45\textwidth]{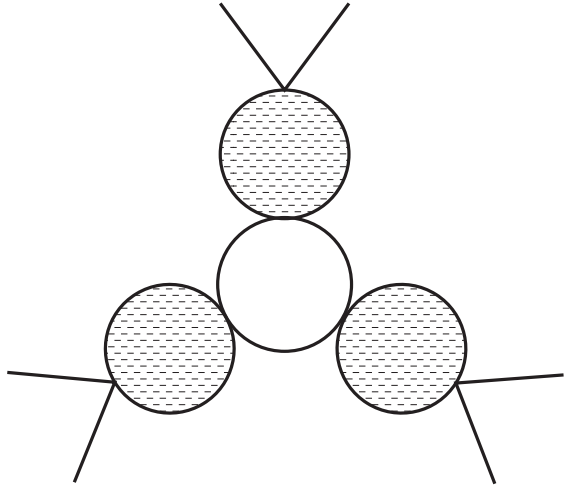}
    \caption{Contribution to an amplitude including the three-graviton vertex. The shaded circle is defined in Fig.~\ref{fig:scat}.}
        \label{fig:threegrav}
  \end{center}
\end{figure}

The preceding argument was made possible by the fact that we started with a form of the theory in which the composite graviton operator $h_{\mu\nu}$ 
appears in the tree-level action in the same way as in general relativity.   We cannot apply the same reasoning in determining, for 
example, the three-graviton coupling, since we have no tree-level terms involving $h_{\mu\nu}$ that we could match to the cubic part 
of $\sqrt{|g|} \, R$.  The graviton self-interaction vertices necessarily arise via connecting scalar loops in the present model: for example, one would 
extract the three-graviton vertex from the diagram shown in Fig.~\ref{fig:threegrav}.  Since the form of the three-graviton vertex is dictated by general 
covariance, and should not be affected by our gauge fixing choice, we expect to recover the same vertex as in general relativity.  It should be possible
to demonstrate this via an explicit evaluation of the amplitude given in Fig.~(\ref{fig:threegrav}).  This calculation is more tedious than the one
discussed in Sec.~\ref{sec:pole} and in the Appendix, and will be included with a more general study of loop corrections in Ref.~\cite{nextpaper}.
   
\section{Discussion and Conclusions}
\label{sec:Conclusions}

We have demonstrated how a constraint of vanishing energy-momentum tensor can lead to a metric-independent theory in which quantum gravity emerges as a nonperturbative artifact of regularization-scale physics. Vanishing of the energy-momentum tensor is closely related to metric-independence of the action, which is desirable for a background-independent description of quantum gravity. 
As an example, we have constructed a scalar field theory with vanishing energy-momentum tensor that has a perturbative low-energy description including scalar fields with a potential. In a large-$N$ limit, we have explicitly demonstrated that 
scattering of scalar particles includes a massless spin-2 pole, corresponding to exchange of a massless composite graviton that couples to matter as in Einstein gravity. The gravitational coupling is determined by the short-distance regularization of the theory, which   for definiteness in our calculations we have defined by dimensional regularization. We conclude by commenting on a number of issues related to the interpretation of these results.

\begin{itemize}
\item
Although our analysis of the theory was done in the large-$N$ limit for simplicity,  we expect that the existence of the massless spin-two state does not depend on  this limit, due to the general covariance of the theory. Furthermore, although  we have not yet computed the gravitational self couplings in the effective theory, we expect those couplings to agree with general relativity up to corrections suppressed by the regularization scale.  These issues deserve further investigation.  


\item
It is worth commenting on renormalizability in this approach. Unless Einstein gravity is asymptotically safe \cite{Weinberg:1976xy}, quantization of the Einstein-Hilbert action is nonrenormalizable: arbitrarily high-dimension operators must be included as corrections to the action in order that correlation functions remain finite as the cutoff tends to infinity, and there is no unique choice of the coefficients of the higher-dimension operators. However, if gravity emerges as a nonperturbative regularization-scale artifact, then the Planck scale is linked to the regularization scale, and higher-dimension operators may remain suppressed by the regularization scale. Hence, renormalizability  is not required by emergent gravity models in the usual sense. However, whether or not  functional integrals in the theory defined by Eq.~(\ref{eq:S}) are well defined nonperturbatively remains an open question.  Similarly, the consequences of introducing additional field dependence in the metric-independent action requires further exploration. 

\item
As a quantum theory of gravity, analysis of the theory at short distances requires that the regularization procedure be treated as physical. For  dimensional regularization, this implies that  functional integrals are to be analytically continued in the number of spacetime dimensions, which is then to be fixed at some $D=4-\epsilon$. The theory determines correlation functions at arbitrarily short distances, but due to the analytic continuation in $D$ those correlators are not expected to satisfy the usual axioms of 3+1-dimensional relativistic quantum field theory; the same could be said for regularization by  Pauli-Villars fields with ``wrong'' spin-statistics properties. 

If we allow for the possibility that short-distance physics violates one or another property usually held dear, such as unitarity of the scattering matrix, then we have an avenue by which to address 
questions related to the singularities that plague classical Einstein gravity. One possibility is that  spacetime remains continuous at arbitrarily short distances, and correlation functions define the physical 
content of the theory, but the particle-like description breaks down at short distances and correlators lose their interpretation in terms of scattering amplitudes. In order to address  issues in quantum 
gravity related to spacetime singularities and spacetimes with horizons, such as the black-hole firewall puzzle  \cite{Almheiri:2012rt}, we need to move beyond the perturbative analysis of this paper and consider background field configurations for which the composite metric is far from the Minkowski metric.

\item
The cosmological constant problem is not immediately resolved in this theory. The composite graviton state couples to the energy-momentum tensor in the same way as the sum of diagrams in Fig.~\ref{fig:interp}. Closing the external lines corresponds to the coupling of the metric fluctuation to the vacuum energy, and leads to the same instability as in general relativity. Hence, even though we showed that the static gauge-fixing condition was consistent with a classical perturbative expansion about a flat composite spacetime,  the requirement of vanishing cosmological constant is necessary for perturbative stability of the flat spacetime. 

\item
It is necessary to understand the coupling of gravity to the Standard Model in this approach. Coupling the Standard Model to an auxiliary vielbein, the vanishing energy-momentum-tensor constraint might determine the vielbein in terms of Standard-Model fields, perhaps similar to the description in Ref.~\cite{Amati:1981tu}. We would again expect the composite vielbein to give rise to  emergent gravitational interactions. The clock and ruler fields $X^I$ might be included in a complete theory, as well as additional scalar fields $\phi^a$.   We note that the $\phi^a$ masses are arbitrary and may be taken large enough in such a scenario to evade potential phenomenological bounds.

\item
Finally, we note  that the Dirac-Born-Infeld action including the gauge field \cite{dbi},
\begin{equation}
S_{{\rm DBI}}\propto\int d^Dx\,\sqrt{\left|\det\left(\partial_\mu{\mathbf X}\cdot\partial_\nu{\mathbf X}+2\pi\alpha'F_{\mu\nu}\right)\right|},
\end{equation}
which describes the dynamics of bosonic ${\rm D}_{p{=}D-1}$ branes (with the Ramond fluxes, the dilaton and the B-field turned off for simplicity), is also metric independent and diffeomorphism invariant. Quite manifestly it is similar to our starting point (\ref{eq:S}). The $X$'s and $\phi$'s in (\ref{eq:S}) 
would be interpreted respectively as coordinates along the D-brane and  transverse to the D-brane. In this sense we would be looking at a ${\rm D}_{p{=}D-1}$ brane embedded in a $D+N$ dimensional spacetime, with a flat bulk Minkowski metric. A difference between our model and the DBI action is 
that in the latter the transverse scalars are massless, since they are Goldstone bosons for the broken translational symmetry, whereas in our theory the masses of the $\phi$'s are nonvanishing. It is tempting to speculate that taking to zero the scalar mass $m$ will not change the conclusion that  generic covariant regulators will generate emergent $D$-dimensional gravity.  With  the string scale playing the role of a regularization scale in the theory, we would therefore expect D-branes to support brane-localized gravity. Thus, string theory appears to 
provide another alternative to the realization of the brane-induced gravity scenario of Dvali, Gabadadze and Porrati~\cite{Dvali:2000hr}, where dynamical bulk gravity gets localized on the brane\footnote{In another related work, \cite{Cheung} computed the open string partition function in the presence of 
a potential that localized the open string endpoints on a D-brane, whose embedding was specified in terms of some transverse scalars $Y^i$. The resulting effective action, as a function of the scalars $Y^i$ which were non-dynamical, was expressed  in terms of the induced metric $g_{\mu\nu}=\eta_{\mu\nu}+\partial_\mu Y^i \partial_\nu Y_i$, and contained a volume term, and higher derivative terms which involved the extrinsic curvature of the brane.  One of these terms was the Einstein-Hilbert action for the induced metric, leading to a picture of open-string induced gravity on the brane.}. 
This possibility was also noted in \cite{Akama:2013tua}.

\end{itemize}
\begin{acknowledgments}  
The work of CC and JE was supported by the NSF under Grant PHY-1519644.  The work of DV was supported in part by the DOE grant DE-SC0007894. 
DV would also like to acknowledge the hospitality of William and Mary Physics Department during the completion of this paper.
\end{acknowledgments}

\begin{appendix}
\section{Integrals}\label{app1}

In this section of the appendix we list the integrals used in the evaluation of the kernel defined in Section \ref{sec:pole}.

We work in the limit when $q^2\ll m^2$, to order ${\cal O}((q^2/m^2)^2)$.
Furthermore, we work here in Euclidean signature.
First, we evaluate
\bea
&&\int d^D l\frac{1}{((l-\tfrac 12 q)^2+m^2)((l+\tfrac 12 q)^2+m^2)}
\nn\\
&&=\int
d^D l\int_0^1 dx \frac{1}{[x((l-\tfrac 12 q)^2+m^2)+(1-x)((l+\tfrac 12 q)^2+m^2)]^2}\nn\\
&&=\int d^D l\int_0^1 dx \frac{1}{\bigg( (l+\tfrac 12 (1-2x) q)^2+m^2+q^2 x(1-x)\bigg)^2}\nn
\eea
\bea
&&=\int d^D l \int_0^1 dx  \int_0^\infty dt \,t e^{-t( l^2 +m^2+x(1-x)q^2)}\nn\\
&&=  (\pi)^{D/2}\int_0^1 dx\int_0^\infty dt \,t^{1-D/2} e^{-t(m^2+x(1-x)q^2)}\nn\\
&&=(\pi)^{D/2} \Gamma(2-\tfrac D2) \int_0^1 dx (m^2+x(1-x)q^2)^{D/2-2} \,\,\,.
\eea
The last integral can be evaluated in terms of an incomplete beta function and subsequently expanded in $q^2\ll m^2$, or we can direcly expand the integrand in $q^2/m^2$ and then do the $x$ integral. The result is the same. To order $(q^2/m^2)^2$ we find
\bea
\int d^D l\frac{1}{((l-\tfrac 12 q)^2+m^2)((l+\tfrac 12 q)^2+m^2)}\simeq
(m^2 \pi)^{D/2} \frac{\Gamma(2-\tfrac D2)}{m^4}
\bigg(1+\frac {D-4}{12} \frac{q^2}{m^2}\bigg) \,\,\,.
\eea

Similarly,
\bea
\int d^D l\frac{l_{\alpha}l_{\beta}}{((l-\tfrac 12 q)^2+m^2)((l+\tfrac 12 q)^2+m^2)}
&\simeq& (m^2 \pi)^{D/2}\frac{\Gamma(1-\tfrac D2)}{m^2}\bigg[\delta_{\alpha\beta}
\bigg(\tfrac 12+\frac{D-2}{24}\frac{q^2}{m^2}\bigg)\nn\\
&&+\frac{(1-\tfrac D2)}{12}\frac{q_{\alpha} q_{\beta}}{m^2}\bigg] \,\,\,,
\eea
and
\bea
&&\!\!\!\!\!\!\!\!\!\!\!\!\!\!\!\int d^D l\frac{l_{\alpha}l_{\beta}l_{\gamma}l_{\delta}}{((l-\tfrac 12 q)^2+m^2)((l+\tfrac 12 q)^2+m^2)}
\simeq  (m^2 \pi)^{D/2}\Gamma({-}\tfrac D2)\bigg[
\frac{\delta_{\alpha\beta}\delta_{\gamma\delta}+\delta_{\alpha\gamma}\delta_{\beta\delta}+\delta_{\alpha\delta}\delta_{\beta\gamma}}4\bigg(1+\tfrac D{12}\frac {q^2}{m^2}\bigg)\nn\\
&&\!\!\!\!\!\!\!\!\!\!\!\!\!\!\!\qquad-\tfrac{D}{48}(\frac{q_\alpha q_\beta}{m^2} \delta_{\gamma\delta}+\frac{q_\alpha q_\gamma}{m^2} \delta_{\beta\delta}+\frac{q_\alpha q_\delta}{m^2} \delta_{\beta\gamma}+\frac{q_{\beta}q_\gamma}{m^2}\delta_{\alpha\delta}+\frac{q_\beta q_\delta}{m^2} \delta_{\alpha \gamma}+\frac{q_\gamma q_\delta}{m^2} \delta_{\alpha\beta})\bigg] \,\,\,.
\eea

\section{Calculation of the kernel}

In Section \ref{sec:pole} we expressed the kernel in terms of the following tensor:
\bea
E_{\alpha\beta}(\tfrac 12 q-l, \tfrac 12 q+l)\equiv 
-\tfrac 12 q_{\alpha} q_\beta + 2 l_\alpha l_\beta + \eta_{\alpha\beta} (\tfrac 14 q^2 -l^2+m^2) \,\,\,,
\eea
where for the sake of clarity we mention that the parentheses on the left hand side denote the argument of the tensor. 

Switching back to the mostly minus Minkowski metric used in the paper, the integral we want to evaluate to order $(q^2/m^2)^2$ is
\bea
{\cal I}_{\alpha\beta|\gamma\delta}\equiv \int d^D l \frac{E_{\alpha\beta}(\tfrac 12 q-l, \tfrac 12 q+l) E_{\gamma\delta}(\tfrac 12 q-l, \tfrac 12 q+l)}{((l-\tfrac 12 q)^2-m^2)((l+\tfrac 12 q)^2-m^2)} \,\,\,.
\eea
This is related to the kernel $K$ as follows (\ref{kern1}):
\be
{K^{\mu\nu}}_{\gamma\delta}=-\frac{iN}{4V_0(2\pi)^D} \bigg((\tfrac D2 -1)\eta^{\mu\alpha}\eta^{\nu\beta} -
\tfrac 12 \eta^{\mu\nu}\eta^{\alpha\beta}\bigg) {\cal I}_{\alpha\beta|\gamma\delta} \,\,\,.
\ee
To this end we employ the integrals computed in Appendix \ref{app1}, and find
\bea
{\cal I}_{\alpha\beta|\gamma\delta}&\simeq&i (m^2 \pi)^{D/2}\Gamma({-}\tfrac D2)\bigg[
-\eta_{\alpha\beta} \eta_{\gamma\delta}+\eta_{\alpha\gamma}\eta_{\beta\delta}+
\eta_{\beta\gamma}\eta_{\alpha\delta}\nn\\
&-&\tfrac{D}4\bigg(\frac{q_\gamma q_\delta}{m^2}\eta_{\alpha\beta}+
\frac{q_\alpha q_\beta}{m^2}\eta_{\gamma\delta}\bigg)
\nn\\
&+&\tfrac{D}{12}\bigg(\frac{q_\alpha q_\beta}{m^2} \eta_{\gamma\delta}
+\frac{q_\gamma q_\delta}{m^2} \eta_{\alpha\beta}+
\frac{q_{\alpha}q_\gamma}{m^2} \eta_{\beta\delta}+
\frac{q_\alpha q_\delta}{m^2}\eta_{\beta\gamma}+
\frac{q_\beta q_\gamma}{m^2}\eta_{\alpha\delta}+
\frac{q_\beta q_\delta}{m^2} \eta_{\alpha\gamma}\bigg)\nn\\
&-& \tfrac{D}{12}(\eta_{\alpha\beta}\eta_{\gamma\delta}+\eta_{\alpha\gamma}\eta_{\beta\delta}+\eta_{\alpha\delta}\eta_{\beta\gamma})\frac{q^2}{m^2}+\tfrac D4\eta_{\alpha\beta}\eta_{\gamma\delta}\frac{q^2}{m^2}\bigg] \,\,\,.
\eea

The kernel becomes
\bea
\!\!\!\!\!\!{K^{\mu\nu}}_{\gamma\delta}(q)&\simeq&\frac{N}{4V_0(2\pi)^D} (m^2\pi)^{D/2}(\tfrac D2-1)\Gamma({-}\tfrac D2) \bigg[(\delta^\mu_{\gamma}\delta^\nu_{\delta}+\delta^\mu_{\delta}\delta^\nu_{\gamma} )\bigg(1-\tfrac{D}{12} \frac {q^2}{m^2}\bigg)\nn\\
&+&\tfrac{D}{12}\bigg(\frac{q^\mu q_\gamma}{m^2}\delta^\nu_\delta+\frac{q^\mu q_\delta}{m^2}\delta^\nu_\gamma+\frac{q^\nu q_\gamma}{m^2}\delta^\mu_\delta+\frac{q^\nu q_\delta}{m^2}\delta^\mu_\gamma\bigg)-\tfrac{D}{6}\frac{q^\mu q^\nu}{m^2}\eta_{\gamma\delta}\bigg] \,\,\,.\label{kernel}
\eea
In the main text we drop the $q^{\mu,\nu}=(p_1+p_2)^{\mu,\nu}$ dependent terms from the kernel, since for the purpose of our calculation we need to contract the kernel with the external line factor $E_{\mu\nu}(p_1, p_2)$ which satisfies (\ref{transv}) $q^\mu E_{\mu\nu}(p_1, p_2)=0$.

\section{The Cosmological Constant}\label{App:CC}

If the counterterm $c_2$ in Eq.~(\ref{eq:dvnew}) is not tuned as described in Section~\ref{sec:pole}, so that $c_2$ is replaced with $c_2+\Delta c_2$, then the effective energy-momentum tensor that couples to the composite graviton becomes \begin{equation}
{\cal T}_{\mu\nu}=\sum_{a=1}^N\bigg[:\partial^\mu \phi^a \partial^\nu \phi^a: - \eta^{\mu\nu} \left(\frac{1}{2} :\partial^\alpha \phi^a \partial_\alpha \phi^a: - \frac{1}{2} m^2 :\phi^a \phi^a: +\Delta c_2\right)\bigg].
\end{equation}
We choose to leave the gauge choice Eq.~(\ref{eq:staticgauge2}) unchanged; otherwise we would  need to take into account the associated wavefunction renormalization of the fields $\phi^a$ in this analysis, but would be led to the same conclusions.

\begin{figure}[t]
  \begin{center}
    \includegraphics[width=0.1\textwidth]{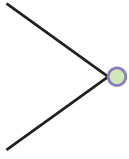}
    \caption{The $\Delta c_2$ vertex.}
        \label{fig:Dc2-vertex}
  \end{center}
\end{figure}

We assume that $\Delta c_2\ll V_0$. In the coupling ${\cal L}_{int}$ of Eq.~(\ref{eq:lint}), we have the following additional interaction, at the same order in $1/V_0$ to which we have been working:
\begin{equation}
{\cal L}_{\Delta c_2}=-\frac{\Delta c_2}{2V_0}\eta^{\mu\nu}{\cal T}_{\mu\nu}.
\label{eq:Dc2int} \end{equation}
The new vertex corresponding to Eq.~(\ref{eq:Dc2int}) is illustrated in Fig.~\ref{fig:Dc2-vertex}. 

\begin{figure}[t]
  \begin{center}
    \includegraphics[width=0.5\textwidth]{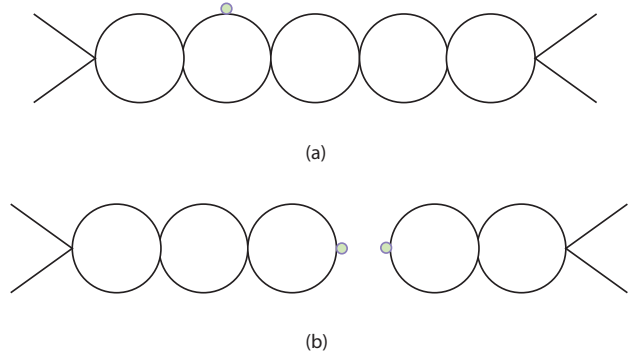}
    \caption{(a) Insertions of the $\Delta c_2$ vertex in scattering diagrams are higher order in $1/V_0$. (b) 
    Diagrams like Fig.~\ref{fig:Dc2-insertions}b are disconnected and do not contribute to scattering
    amplitudes. }
        \label{fig:Dc2-insertions}
  \end{center}
\end{figure}
\begin{figure}[t]
  \begin{center}
    \includegraphics[width=0.5\textwidth]{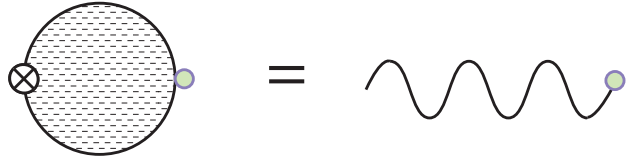}
    \caption{The $\Delta c_2$ vertex sources the composite metric as a cosmological constant.}
        \label{fig:CosmoConst}
  \end{center}
\end{figure}

As illustrated in Fig.~\ref{fig:Dc2-insertions},  $\Delta c_2$ does not contribute to scattering amplitudes at leading order in $1/N$. Hence, to this order the argument presented in Sec.~\ref{sec:pole} for the massless graviton pole remains unchanged. However,  the constant  $\Delta c_2$  in the scalar potential acts as a  source for the composite metric $h_{\mu\nu}$, as per the diagram in Fig.~\ref{fig:CosmoConst}. This source corresponds to the tadpole instability of the Minkowski-space vacuum in the presence of the cosmological constant. Hence, the cosmological constant problem appears to remain in this emergent gravity scenario.
\end{appendix}


\begin{thebibliography}{99}
\bibitem{Bjorken:1963vg} 
  J.~D.~Bjorken,
  ``A Dynamical origin for the electromagnetic field,''
  Annals Phys.\  {\bf 24}, 174 (1963).
  
\bibitem{Eguchi:1976iz} 
  T.~Eguchi,
  ``A New Approach to Collective Phenomena in Superconductivity Models,''
  Phys.\ Rev.\ D {\bf 14}, 2755 (1976).

\bibitem{Amati:1974rm} 
  D.~Amati and M.~Testa,
  ``Quark imprisonment as the origin of strong interactions,''
  Phys.\ Lett.\ B {\bf 48}, 227 (1974).

\bibitem{Rajasekaran:1977td} 
  G.~Rajasekaran and V.~Srinivasan,
  ``Generation of Gluons from Quark Confinement,''
  Pramana {\bf 10}, 33 (1978).
  
\bibitem{Suzuki:2016aqj} 
  M.~Suzuki,
  ``Composite gauge-bosons made of fermions,''
  Phys.\ Rev.\ D {\bf 94}, no. 2, 025010 (2016)
  [arXiv:1603.07670 [hep-th]].

\bibitem{Sindoni:2011ej} 
  L.~Sindoni,
  ``Emergent Models for Gravity: an Overview of Microscopic Models,''
  SIGMA {\bf 8}, 027 (2012)
  [arXiv:1110.0686 [gr-qc]].

\bibitem{Sakharov:1967pk} 
  A.~D.~Sakharov,
  ``Vacuum quantum fluctuations in curved space and the theory of gravitation,''
  Sov.\ Phys.\ Dokl.\  {\bf 12}, 1040 (1968)
  [Dokl.\ Akad.\ Nauk Ser.\ Fiz.\  {\bf 177}, 70 (1967)]
  [Sov.\ Phys.\ Usp.\  {\bf 34}, 394 (1991)]
  [Gen.\ Rel.\ Grav.\  {\bf 32}, 365 (2000)].

\bibitem{Terazawa:1976xx} 
  H.~Terazawa, K.~Akama and Y.~Chikashige,
  ``Unified Model of the Nambu-Jona-Lasinio Type for All Elementary Particle Forces,''
  Phys.\ Rev.\ D {\bf 15}, 480 (1977).
  
  \bibitem{Akama:1977hr} 
  K.~Akama, Y.~Chikashige, T.~Matsuki and H.~Terazawa,
  ``Gravity and Electromagnetism as Collective Phenomena: A Derivation of Einstein's General Relativity,''
  Prog.\ Theor.\ Phys.\  {\bf 60}, 868 (1978).
  
\bibitem{Amati:1981rf} 
  D.~Amati and G.~Veneziano,
  ``Metric From Matter,''
  Phys.\ Lett.\ B {\bf 105}, 358 (1981).

  
 \bibitem{Amati:1981tu} 
  D.~Amati and G.~Veneziano,
  ``A Unified Gauge and Gravity Theory With Only Matter Fields,''
  Nucl.\ Phys.\ B {\bf 204}, 451 (1982).

\bibitem{Visser:2002ew} 
  M.~Visser,
  ``Sakharov's induced gravity: A Modern perspective,''
  Mod.\ Phys.\ Lett.\ A {\bf 17}, 977 (2002)
  [gr-qc/0204062].

\bibitem{Polyakov:1981rd} 
  A.~M.~Polyakov,
  ``Quantum Geometry of Bosonic Strings,''
  Phys.\ Lett.\ B {\bf 103}, 207 (1981).

  
 \bibitem{DeWitt:1967yk} 
  B.~S.~DeWitt,
  ``Quantum Theory of Gravity. 1. The Canonical Theory,''
  Phys.\ Rev.\  {\bf 160}, 1113 (1967).

\bibitem{Akama:1978pg} 
  K.~Akama,
  ``An Attempt at Pregeometry: Gravity With Composite Metric,''
  Prog.\ Theor.\ Phys.\  {\bf 60}, 1900 (1978).
  
  \bibitem{Weinberg:1964ew} 
  S.~Weinberg,
  ``Photons and Gravitons in s Matrix Theory: Derivation of Charge Conservation and Equality of Gravitational and Inertial Mass,''
  Phys.\ Rev.\  {\bf 135}, B1049 (1964).
  
  \bibitem{Weinberg:1980kq} 
  S.~Weinberg and E.~Witten,
  ``Limits on Massless Particles,''
  Phys.\ Lett.\ B {\bf 96}, 59 (1980).
  
  \bibitem{Porrati:2008rm} 
  M.~Porrati,
  ``Universal Limits on Massless High-Spin Particles,''
  Phys.\ Rev.\ D {\bf 78}, 065016 (2008)
  [arXiv:0804.4672 [hep-th]].

\bibitem{Gibbons:2004dz} 
  G.~W.~Gibbons and A.~Ishibashi,
  ``Topology and signature changes in brane worlds,''
  Class.\ Quant.\ Grav.\  {\bf 21}, 2919 (2004)
  [hep-th/0402024].

\bibitem{Page:1983uc} 
  D.~N.~Page and W.~K.~Wootters,
  ``Evolution Without Evolution: Dynamics Described By Stationary Observables,''
  Phys.\ Rev.\ D {\bf 27}, 2885 (1983).

\bibitem{nextpaper} C.~D.~Carone and D.~Vaman, in progress.
  
  \bibitem{Weinberg:1976xy} 
  S.~Weinberg,
  ``Critical Phenomena for Field Theorists,''
  HUTP-76-160, Erice Subnucl. Phys. (Lectures presented at Int. School of Subnuclear Physics); 
  
\bibitem{Almheiri:2012rt} 
  A.~Almheiri, D.~Marolf, J.~Polchinski and J.~Sully,
  ``Black Holes: Complementarity or Firewalls?,''
  JHEP {\bf 1302}, 062 (2013)
  [arXiv:1207.3123 [hep-th]].

\bibitem{dbi}
J.~Polchinski,  ``Dirichlet Branes and Ramond-Ramond Charges,'' Phys. Rev. Lett., {\bf 75}, 4724 (1995).  [arXiv:hep-th/9510017];

 E.~S.~Fradkin, and A.~A.~Tseytlin, ``Non-linear Electrodynamics from Quantized Strings,'' Phys. Lett. B, {\bf 163}, 123 (1985);


R.~G.~Leigh, ``Dirac-Born-Infeld Action from Dirichlet-Model,''  Mod. Phys. Lett. A, {\bf 4}, 2767 (1989);

M.~Born and L.~Infeld, ``Foundations of the new field theory,'' Proc. R. Soc. London, Ser. A, {\bf 144}, 425 (1934). 

  \bibitem{Dvali:2000hr} 
  G.~R.~Dvali, G.~Gabadadze and M.~Porrati,
  ``4-D gravity on a brane in 5-D Minkowski space,''
  Phys.\ Lett.\ B {\bf 485}, 208 (2000)
  [hep-th/0005016].

\bibitem{Cheung}
  Y.~K.~E.~Cheung, M.~Laidlaw and K.~Savvidy,
  ``Open string gravity?,''
  JHEP {\bf 0412}, 028 (2004)
  [hep-th/0406245].

  \bibitem{Akama:2013tua} 
  K.~Akama and T.~Hattori,
  ``Dynamical Foundations of the Brane Induced Gravity,''
  Class.\ Quant.\ Grav.\  {\bf 30}, 205002 (2013)
  [arXiv:1309.3090 [gr-qc]].






\end{thebibliography}
\end{document}